# Automated, real-time hospital ICU emergency signaling: A field-level implementation

Nazifa M Shemonti, Shifat Uddin, Saifur Rahman, Tarem Ahmed and Mohammad Faisal Uddin
*Department of Computer Science and Engineering*
*Independent University, Bangladesh*
Dhaka, Bangladesh
{nazifa, shifat.uddin, saifur, tarem, faisal}@iub.edu.bd

*Abstract*—Contemporary patient surveillance systems have streamlined central surveillance into the electronic health record interface. They are able to process the sheer volume of patient data by adopting machine learning approaches. However, these systems are not suitable for implementation in many hospitals, mostly in developing countries, with limited human, financial, and technological resources. Through conducting thorough research on intensive care facilities, we designed a novel central patient monitoring system and in this paper, we describe the working prototype of our system. The proposed prototype comprises of inexpensive peripherals and simplistic user interface. Our central patient monitoring system implements Kernel-based Online Anomaly Detection (KOAD) algorithm for emergency event signaling. By evaluating continuous patient data, we show that the system is able to detect critical events in real-time reliably and has low false alarm rate.

## I. INTRODUCTION

Bedside patient monitoring devices in the hospital Intensive Care Unit (ICU) generate massive quantities of patient surveillance data. Medical equipment supply companies have shifted toward applying machine learning to process the data. Prominent central monitoring frameworks e.g., Philips IntelliVue Information Center iX, Cambridge Enterprise ICM+, OBS Medical Visensia®, and Excel Medical Wave are already deployed in large healthcare systems. Philips IntelliVue Information Center iX keeps continuous patient record, supports third-party ICU devices configuration, and implements networking best practices [1]. Analog and digital monitoring of brain, autoregulation calculation, and event flagging with fixed timestamp and username are built-in attributes of Cambridge Enterprise ICM+ [2]. The patient surveillance platform Wave, developed by Excel Medical, features Visencia® technology. The clinical decision support platform can display patient data from supported monitoring units in addition to calculate early-warning index by analyzing patterns of up to five vital signs [3]. Automated patient monitoring, real time data analysis and intelligent alarming features have increasingly become the industry standard for ICU in hospitals.

Although these frameworks are available commercially, they are not globally ubiquitous within healthcare systems especially in local hospital ICUs in developing countries with limited financial and technological resources.

To that effect, we propose Kernel-based Online Anomaly Detection (KOAD) algorithm [4] implemented automated hospital ICU signaling system. The system integrates KOAD, a lightweight anomaly detection algorithm in terms of computational and memory resources, with cost effective modular components. Basically, a transceiver unit extracts patient data from the Patient Monitoring Unit (PMU) and transmits the numerical data stream to central monitoring unit. The KOAD patient surveillance application evaluates up to five real-time patient statuses simultaneously. Then instantaneously sounds an alarm when anomaly is detected. The central monitor displays a straightforward output that can be easily interpreted by ICU staff. The system does not involve unsustainable costs for staff training and equipment maintenance. Hence, our proposed system is amenable for ICUs in many developing countries.

The rest of this paper is organized as follows: Section II provides background information on KOAD. The related works and previous work on anomaly detection using KOAD are discussed in section III. The approaches consulted in designing the prototype is detailed in section IV. Section V presents the implementation of the prototype. The performance analysis of the prototype is shown in section VI. Analysis of the system and future avenues of work on the research are provided in section VII.

## II. KERNEL-BASED ONLINE ANOMALY DETECTION ALGORITHM

Ahmed et al. [4], [5] presented KOAD based on kernel mapping functions, used by kernel machines to produce non-linear, non-parametric, efficient, and online machine learning algorithms. Termed the "kernel trick", the idea is that using a suitable kernel function, the input data can be mapped onto a higher dimensional Hilbert feature space [6].

KOAD operates at each timestep $t$ on a input vector $x_t$. It begins by evaluating the projection error $\delta_t$ of the measurement vector onto the current dictionary:

$$\delta_t = \min_a \left\| \sum_{j=1}^{m} \alpha_j \varphi(\tilde{x}^j) - \varphi(x^t) \right\|^2 < \nu$$

If the incoming vector satisfies this minimum threshold or in other words, if the vector is adding new information to the predictor, this measurement vector will be added to the feature space. Hence the dimension of the feature space does not increase out of bounds and there is no requirement for unreasonable memory and computational requirements.

This error measure $\delta_t$ is then compared with two thresholds $\nu_1$ and $\nu_2$, where $\nu_1 < \nu_2$.

If $\delta_t < \nu_1$, KOAD infers that the input data is sufficiently linearly dependent on the dictionary, and thus represents normal behavior.

If $\delta_t > \nu_2$, KOAD concludes that is far away from the realm of normality and consequently raises a "Red1" to immediately signal an anomaly.

Where $\nu_1 < \delta_t < \nu_2$, KOAD infers that is sufficiently linearly independent from the dictionary to be considered an unusual event. It may indeed be an anomaly, or it may represent an expansion or migration of the space of normality itself. In this case, KOAD does the following: it immediately raises an "Orange" alarm, then keeps track of the contribution of the relevant input vector in explaining subsequent arrivals for a further $l$ time step, and then finally at time $t + l$ resolves into either "Green" meaning normal observation or a "Red2" alarm meaning anomalous observation.

The properties of KOAD are: incorporate two thresholds of approximate linear independence; implement exponential forgetting to gradually reduce the importance of past observations; and keep the basis set to dynamically remain current by employing the deletion of previous dictionary members.

The storage complexity of the KOAD depends on the maximum dimensions of the variables and size of the dictionary computational complexity of the KOAD is $O(m^2)$, for every usually much less than the number of input vectors). The from dictionary. Overall, the time-to-detection of anomalies is very fast for KOAD.

### III. LITERATURE REVIEW

Ahmed et al. showed in [7] that, by using Gaussian kernel function, KOAD is able to detect emergency events. They collected real data manually recorded in tabular sheets of 3 patients: 100 rows of 37 vital statistics logged at 1-minute intervals with critical values identified by a cardiac surgeon. They demonstrated that, for KOAD, the emergency event detection rate was high with the least number of false alarms. The implementation and success of KOAD in multivariate data anomaly detection are found in the series of works by Tarem et al. in [4], [8], [9]. They illustrate real-time applications of KOAD including malicious attack detection in network flows in high-speed backbones, intruder detection in automated visual surveillance, Distributed Denial of Service and Flash event detection in internet based service or application etc.

Kernel methods have been used in biology to predict the formation of disulphide bridges in protein [10]. In this study, the authors aimed at developing a swift technique that prevails old approaches and yield higher percentage of exact connectivity pattern prediction. KRLS was used to build a neural network termed Cerebellar Model Articulation Controller (CMAC).

CMAC is able to model the part of the brain responsible for fine muscle control in animals and with the aid of sparsification the network takes lesser time [11].

### IV. METHODOLOGY

In order to construct an effective patient surveillance system, we took a step by step approach to familiarize with intensive care facilities. First, we considered workforce data published for hospital ICU monitoring in Bangladesh. Faruq et al. deduced that there is a shortage of intensive care staff in local hospitals and the existing staff are insufficiently trained to operate complex devices [12]. The authors in [13] classified heavy workload, lack of communication, inadequate supervision as risk factors; and recommended redesign of medical equipment toward reducing injuries and death as a result of care related causes. These concordant studies highlighted key surveillance issues.

Second, we surveyed 5 hospital ICUs in Bangladesh located at Square Hospital, Dhaka Medical College Hospital, Apollo Hospital, Islami Bank Hospital, Euro-Bangla Heart Hospital, Chittagong Medical Center and National Heart Foundation & Research Institute (NHFH&RI). The primary surveys gave us firsthand understanding of ICU setup and models of care in each hospital.

The ICU team at NHFH&RI consists of intensive care nurses, specialist physicians, and readily available healthcare professionals with formal ICU training. The nurses provide nursing care round the clock. Also, they keep manual record of hemodynamic, ventilation, sustenance, and digestion system vital signals from various bedside devices on tabular sheets. The tabular informs specialist physician of the patient's progress and the physician plans the course of treatment accordingly. Patient care, in essence, is a combination of scheduled visits from intensive care nurses and alert-responded consultations by specialist physicians [14]. However, manual records are unreliable due to having limited function. They can introduce inaccuracy from incomplete entries; they are often difficult to read; and they may be lost or otherwise become inaccessible [15].

Typical bedside PMUs continuously display the physiological data of patients. They sound warnings when the built-in change point detection algorithm detects deviation in any of the parameters. In reality, the devices bleep continuously which contributes to an innate alarm fatigue among the nurses. As presented in [16], proper respiration ensures ample supply of oxygen to tissues; the extent of oxygenation and perfusion of tissues determine blood pressure levels. The existing PMUs fail to address this inherent correlation between the biological signals.

Third, we investigated various PMU vendors and available real-time data extraction methods for each variety. We procured ethical clearance from the Ethical Review Committee (ERC) of NHFH&RI to extract physiological data from their PMUs, namely Philips IntelliVue Patient Monitor MP60, on the basis of protecting the anonymity of our subjects. At NHFH&RI, all the PMUs are networked to an underutilized central monitoring display via LAN switch. The central monitoring unit does not provide any data extraction or printing option and therefore we decided to extract data from each PMU and relay the data to an online server.

Last, we studied texts authored by medical experts. Based on the classification of crucial factors that delineate clinical care and intensive care, we designed the user interface. The factors include: capacity of therapeutic personnel, area design, integration with monitoring technology [17]. The international practice guideline [18] considers the necessity of a unit intensive care patient database. We augmented properties of medical equipment alarming systems mentioned in [19]. This secondary survey of guidelines, clinical protocols, and operational references were useful to integrate our prototype with existing architecture.

## V. Implementation of Automated, Real-time Hospital ICU Emergency Signaling

Our working prototype integrates a data extraction adapter, a transceiver computer, a router, a KOAD computer, and a central display. The MediCollector adapter connects to the MIB/RS232 port of the MP60 via type RJ45 network cable. The adapter links to the COM port of a Raspberry Pi via USB to Serial adapter. A Dell desktop and a D-Link router were installed inside the NHFH&RI ICU area. We established a wireless connection between the Raspberry Pi and the desktop.

The technical specification of each component is as follows:
- Adapter: Medicollector adapter CBL-5010
- Transceiver Computer: Raspberry Pi 3 Model B+ with 1.4GHz quad-core Broadcom BCM2837B0 processor, 1GB LPDDR2 SDRAM, and 16GB SD card
- Router: D-Link DIR-600M 150Mbps Wireless Router
- KOAD Computer: Dell Optiplex with 3.0-4.1GHz 8th Gen Intel Core i5 8500 processor, 4GB DDR4 RAM, and 1TB HDD
- Display: Dell E1916HV 18.5" LED Monitor

The operating systems for transceiver computer and KOAD computer are Raspbian Buster and Windows 10 Pro respectively. VSCapture program is installed on transceiver computer to extract data and store as comma-separated values (CSV) file format.

The KOAD patient surveillance application is currently a python script that runs on KOAD computer. The script executes VSCapture program remotely and prompts to extract data from PMU at 12-second intervals. Once VSCapture begins to store data in csv file on transceiver computer, the script converts each row into a string. Next, the script sends the string to KOAD computer.

In KOAD computer, the string is converted to an array of data and undergoes a validity check. This approach ensures preliminary data cleaning and that only valid rows get inserted into KOAD database. A MATLAB script runs KOAD algorithm on newly arriving data stream and the signals are displayed accordingly on the monitor.

Finally, the system immediately alerts when anomalies are detected. The signals are displayed accordingly on the monitor.

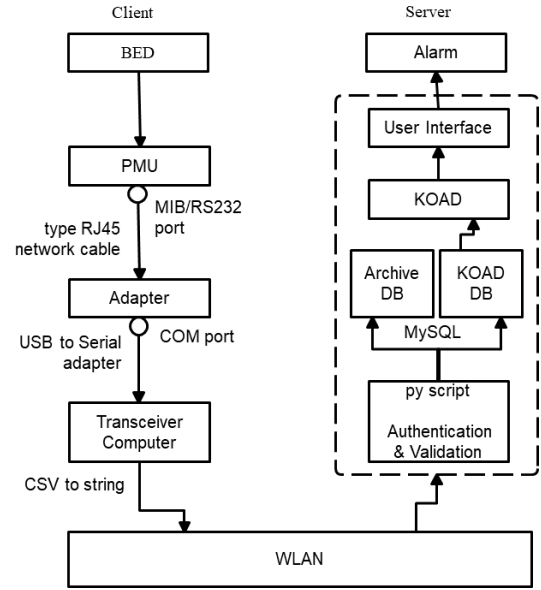

Fig. 1. The connections of the emergency signaling prototype.

Fig. 1 illustrates the underlying data flow of our surveillance system.

### A. Validity Check

Numerical data streams often get interrupted during data extraction from electrocautery interference or occasional patient movement. The validity check removes the affect of interruption and flags archive database entries which contain invalid rows. Each row of data is subjected to the process illustrated in fig. 2.

The process starts by matching the password in the leading array. Then each value in the array with correct password is compared with erroneous characters: null, 0, hyphen, and greater than 10000. On one hand, arrays with no flag are inserted to KOAD database. On the other hand, when numerous consecutive arrays are flagged, a warning is displayed on the central monitor. This process executes until the patient is discharged from the PMU.

### B. Central Surveillance User Interface

The user interface is continuously updated to display the underlying situation of each patient connected to our surveillance system. The display shows upto five patient statuses and a summary of emergency events detected and addressed. The elements in the monitor are demonstrated in fig. 3.

When the system detects immediate and threatening danger which requires prompt attention, the corresponding bed element flashes red. The bed element turns orange when the system detects gradual change that indicates imminent danger.

On the occasion when the signaling system fails to extract valid signals, the bed element alerts the nurses. When this happens, the common console shows a warning sign until the validity check receives reliable data. This method of warning announcement showed to be intuitive for ICU staff.

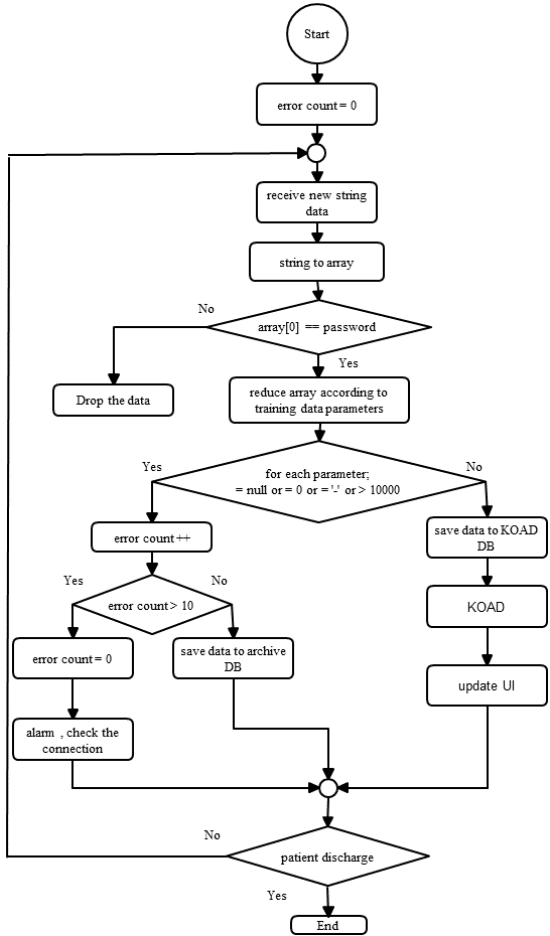

Fig. 2. The validity checking scheme on newly arriving data.

## VI. Experiments

To evaluate the performance of the KOAD patient surveillance application, we monitored 5 patients continuously for 8 hours. VSCapture is able to extract 23 parameters and 40,000 rows of patient data with 89% success rate (∈ 35,600 rows of valid data instances). The success rate is calculated after running the validity check. For training and testing, we omitted time, relative time, premature ventricular contraction, and blood pressure ventricular contraction parameters for irrelevance. We scrambled any identifiers in the patient data to preserve the patients' anonymity. Instead, we allocated randomized bed number to each recorded dataset.

For our experiment, we utilized a supervised cross validation approach in the training period. This method obtains the best pair of values for the thresholds $v_1$ and $v_2$. For the subsidiary parameters, we used default initialization values because the detection result of KOAD algorithm does not depend on the particular setting of these subsidiary parameters.

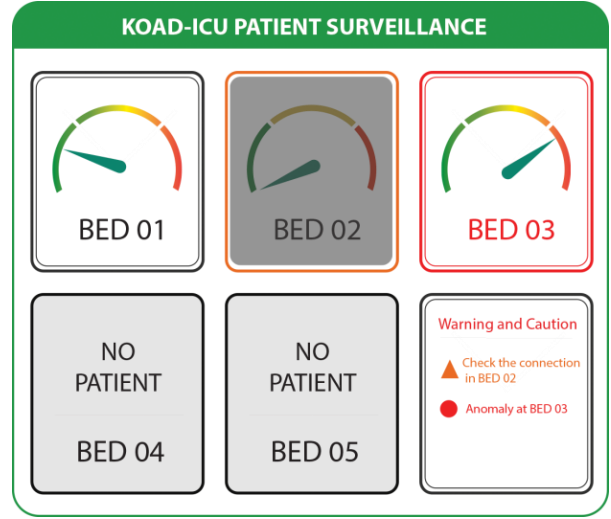

Fig. 3. The central monitoring display.

We ran KOAD in real-time on 300 timesteps with events manually identified by an ICU specialist. Table I depicts the comparable anomaly detection rates for 3 sets of thresholds. The goal was to obtain the best combination of $v_1$ and $v_2$ for a particular patient. Based on the outcome of the training data, we used $v_1 = 0.07$, $v_2 = 0.16$ to detect anomalies in the subsequent 1700 timesteps. We can see in fig. 4 that, $v_1 = 0.07$, $v_2 = 0.16$ correctly identified 27 out of 36 anomalies in the dataset of total 2000 timesteps.

TABLE I
DETECTION OF 9 ANOMALIES WITH KOAD USING VARIOUS SETTINGS OF $v_1$ AND $v_2$

| $v_1$ setting | $v_2$ setting | Detected | Missed | False |
|---|---|---|---|---|
| 0.03 | 0.08 | 5 | 4 | 4 |
| 0.07 | 0.16 | 6 | 3 | 2 |
| 0.11 | 0.24 | 4 | 5 | 6 |

## VII. Discussion and Future Work

We have described the working prototype of KOAD algorithm implemented automated hospital ICU signaling system. The scope of our system does not include inferring whether patient condition has improved. The interface does not offer any solution to malfuctions in operative components.

The prototype is able to extract patient data using inexpensive modular components and detect emergency events based on the data in real time. Patients were not exposed to any risk during the course of our study. Our system does not cause any physical or psychological discomfort, invasion of privacy, or disclosure of damaging information of the subjects.

We intend to further modify the prototype to accommodate some useful features. Currently KOAD patient surveillance application is manually executed and terminated.

The application will be reconfigured so that ICU staff require minimal training to operate the system. The algorithm detects anomalous events with high accuracy and low false alarm rates. Our objective is to improve the algorithm to set optimum values to all thresholds and subsidiary parameters. This automation will decrease alarm fatigue. The monitor has no feedback option at present. We are focused to construct an interactive display that can be customized according to ICU capacity. The existing archive database can be programmed to prepare various summary reports. ICUs will be less dependent on paper records this way and they will have access to accurate digital records as well. By adopting these features into our prototype, the proposed system can be installed as the primary central monitoring framework in local hospitals.

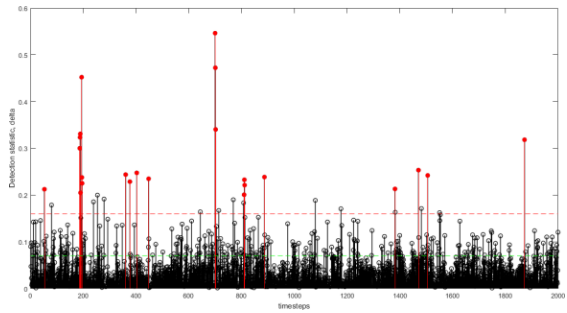

Fig. 4. Sequence of the detection statistic $\delta_t$ with KOAD algorithm run in real time over full set of 14 parameters across 2000 timesteps. Sample experiment with $v_1$=0.07, $v_2$=0.16. Critical events are indicated with red, filled stems.

The KOAD algorithm implemented automated hospital ICU signaling system aims to remove global disparities in ICU patient monitoring. The prototype is suitable for understaffed ICUs with technological constraints. The objective of implementing our prototype is to eliminate the irregularities in patient monitoring. The system can be utilized as an educational tool to research sustainable patient surveillance approaches in hospitals.